\documentclass[aps,prl,twocolumn,superscriptaddress,showpacs,preprintnumbers,amsmath,amssymb]{revtex4}

\def\bellelogo{\vbox to 16mm{
               \vss\hbox to \textwidth{\resizebox{!}{2cm}{
               \includegraphics{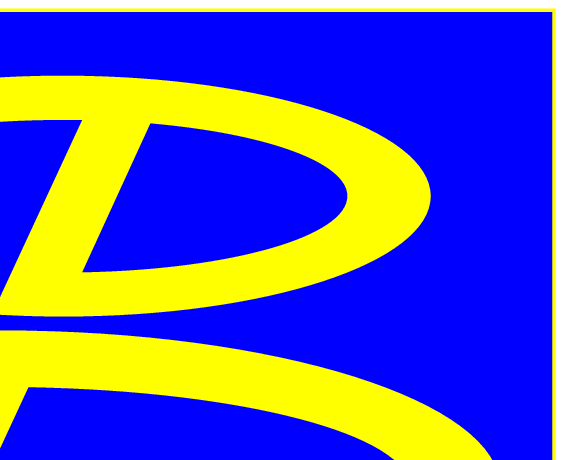}}\hss }}\vspace{-1cm}}
\def\preprintA{\hbox{\hfil KEK Preprint 2005-103}}
\def\preprintB{\hbox{\hfil Belle Preprint 2006-7}}

\usepackage{epsfig}
\usepackage{dcolumn}  
\usepackage{relsize}
\usepackage{psfrag}
\usepackage{tabularx}
\usepackage{color}
\usepackage{graphicx}

\graphicspath{{ps}}

\definecolor{red}{rgb}{1,0,0}
\definecolor{green}{rgb}{0,0.7,0}
\definecolor{violet}{rgb}{0.4,0.0,0.55}
\definecolor{blue}{rgb}{0,0,0.7}
\definecolor{magenta}{rgb}{0.8,0.0,0.5}

\newcommand{\BB}{\ensuremath{B\overline{B}}}
\newcommand{\Bbar}{\ensuremath{\overline{B}}}
\newcommand{\Bbaro}{\ensuremath{\overline{B}{}^0}}

\newcommand{\de}{\ensuremath{\Delta E}}
\newcommand{\mb}{\ensuremath{M_{\mbox{\scriptsize bc}}}}
\newcommand{\etap}{\ensuremath{\eta^\prime}}
 
\newcommand{\BF}{\ensuremath{{\mathcal B}}}

\newcommand{\LK}{\ensuremath{{\cal L}}}

\newcommand{\FD}{\ensuremath{{\cal F}}}
\newcommand{\LR}{\ensuremath{{\cal R}}}
\newcommand{\hel}{\ensuremath{{\cal H}}}
\newcommand{\cost}{\ensuremath{\cos\theta_T}}

\newcommand{\sperp}{\ensuremath{S_\perp}}
\newcommand{\epp}{\ensuremath{\eta' \to \eta \pi^+ \pi^-}}
\newcommand{\erg}{\ensuremath{\eta' \to \rho^0 \gamma}}
\newcommand{\ebeam}{\ensuremath{E_{\mbox{\scriptsize beam}}}}

\newcommand{\cs}{\ensuremath{/c^2}}
\newcommand{\beq}{\begin{eqnarray}}
\newcommand{\eeq}{\end{eqnarray}}
\newcommand{\eff}{\ensuremath{\epsilon}}
 
\newcommand{\ACP}{\ensuremath{A_{CP}}}

\newcommand{\btepk}{\ensuremath{B \to \etap K}}
\newcommand{\bteppi}{\ensuremath{B \to \etap \pi}}
\newcommand{\btepkp}{\ensuremath{B^+ \to \etap K^+}}
\newcommand{\bteppip}{\ensuremath{B^+ \to \etap \pi^+}}
\newcommand{\btepks}{\ensuremath{B^0 \to \etap K^0}}
\newcommand{\bteppio}{\ensuremath{B^0 \to \etap \pi^0}}

\begin{document}

%
%
\title{
  \vbox{\bellelogo \vspace{-1cm}
  \hbox to \textwidth{\rm\normalsize\hss\preprintA}
  \hbox to \textwidth{\rm\normalsize\hss\preprintB}}
  \vspace*{15mm}Evidence for \bteppi{} and improved measurements for
	\btepk}

\affiliation{Budker Institute of Nuclear Physics, Novosibirsk}
\affiliation{Chiba University, Chiba}
\affiliation{Chonnam National University, Kwangju}
\affiliation{University of Cincinnati, Cincinnati, Ohio 45221}
\affiliation{University of Frankfurt, Frankfurt}
\affiliation{University of Hawaii, Honolulu, Hawaii 96822}
\affiliation{High Energy Accelerator Research Organization (KEK), Tsukuba}
\affiliation{Institute of High Energy Physics, Chinese Academy of Sciences, Beijing}
\affiliation{Institute of High Energy Physics, Vienna}
\affiliation{Institute of High Energy Physics, Protvino}
\affiliation{Institute for Theoretical and Experimental Physics, Moscow}
\affiliation{J. Stefan Institute, Ljubljana}
\affiliation{Kanagawa University, Yokohama}
\affiliation{Korea University, Seoul}
\affiliation{Kyungpook National University, Taegu}
\affiliation{Swiss Federal Institute of Technology of Lausanne, EPFL, Lausanne}
\affiliation{University of Maribor, Maribor}
\affiliation{University of Melbourne, Victoria}
\affiliation{Nagoya University, Nagoya}
\affiliation{Nara Women's University, Nara}
\affiliation{National Central University, Chung-li}
\affiliation{National United University, Miao Li}
\affiliation{Department of Physics, National Taiwan University, Taipei}
\affiliation{H. Niewodniczanski Institute of Nuclear Physics, Krakow}
\affiliation{Nippon Dental University, Niigata}
\affiliation{Niigata University, Niigata}
\affiliation{Nova Gorica Polytechnic, Nova Gorica}
\affiliation{Osaka City University, Osaka}
\affiliation{Osaka University, Osaka}
\affiliation{Panjab University, Chandigarh}
\affiliation{Peking University, Beijing}
\affiliation{Princeton University, Princeton, New Jersey 08544}
\affiliation{RIKEN BNL Research Center, Upton, New York 11973}
\affiliation{Saga University, Saga}
\affiliation{University of Science and Technology of China, Hefei}
\affiliation{Seoul National University, Seoul}
\affiliation{Sungkyunkwan University, Suwon}
\affiliation{University of Sydney, Sydney NSW}
\affiliation{Tata Institute of Fundamental Research, Bombay}
\affiliation{Toho University, Funabashi}
\affiliation{Tohoku Gakuin University, Tagajo}
\affiliation{Tohoku University, Sendai}
\affiliation{Department of Physics, University of Tokyo, Tokyo}
\affiliation{Tokyo Institute of Technology, Tokyo}
\affiliation{Tokyo Metropolitan University, Tokyo}
\affiliation{Tokyo University of Agriculture and Technology, Tokyo}
\affiliation{University of Tsukuba, Tsukuba}
\affiliation{Virginia Polytechnic Institute and State University, Blacksburg, Virginia 24061}
\affiliation{Yonsei University, Seoul}
   \author{J.~Sch\"umann}\affiliation{National United University, Miao Li} 
   \author{C.~H.~Wang}\affiliation{National United University, Miao Li} 
   \author{K.~Abe}\affiliation{High Energy Accelerator Research Organization (KEK), Tsukuba} 
   \author{I.~Adachi}\affiliation{High Energy Accelerator Research Organization (KEK), Tsukuba} 
   \author{H.~Aihara}\affiliation{Department of Physics, University of Tokyo, Tokyo} 
   \author{D.~Anipko}\affiliation{Budker Institute of Nuclear Physics, Novosibirsk} 
   \author{K.~Arinstein}\affiliation{Budker Institute of Nuclear Physics, Novosibirsk} 
   \author{Y.~Asano}\affiliation{University of Tsukuba, Tsukuba} 
   \author{T.~Aushev}\affiliation{Institute for Theoretical and Experimental Physics, Moscow} 
   \author{A.~M.~Bakich}\affiliation{University of Sydney, Sydney NSW} 
% ...
\collaboration{The Belle Collaboration}

\tighten

\begin{abstract}
We report evidence for exclusive two-body charmless hadronic $B$ meson
decays \bteppi{}, and improved measurements of \btepk{}. The
results are obtained from a data sample of $386 \times 10^6$ \BB{} pairs
collected at the $\Upsilon(4S)$ resonance, with the Belle detector at the KEKB
asymmetric energy $e^+e^-$ collider. We measure 
$\BF(\bteppip)=[ 1.76 ^{+0.67}_{-0.62}$(stat)$^{+0.15}_{-0.14}$(syst)]$ 
\times 10^{-6}$ and
$\BF(\bteppio)=[ 2.79 ^{+1.02}_{-0.96}$(stat)$^{+0.25}_{-0.34}$(syst)]$
\times 10^{-6}$. 
We also report the ratio of $\frac{\BF(\btepkp)}{\BF(\btepks)}
= 1.17 \pm 0.08$(stat)$ \pm 0.03$(syst) and direct $CP$ asymmetries for 
the charged modes.
\end{abstract}

\pacs{11.30.Er,11.30.Hv,13.20.He,13.25Hw,14.40.Nd,14.65.Fy}

\maketitle

\tighten

{\renewcommand{\thefootnote}{\fnsymbol{footnote}}}
\setcounter{footnote}{0}

Information on the two-body charmless hadronic $B$ meson decays
\bteppi{} is very limited at present~\cite{CC}.
Measurements of these decay modes can improve the understanding
of the flavor-singlet penguin amplitude with intermediate
$t$, $c$ and $u$ quarks~\cite{Chiang:2004nm}.
Theoretical predictions for the branching
fractions cover the range (1--17)$ \times 10^{-6}$ and (1--8)$ \times 10^{-6}$ 
for the charged and neutral decays, respectively~\cite{Chiang:2004nm,Fu:2003fy}.
Recently the charged decay was measured by 
BaBar~\cite{Aubert:2005bq}.
In contrast, the channel \btepk{} has been precisely
measured~\cite{Richichi:1999kj,Abe:2001pf,Aubert:2005iy}.
In the Standard Model (SM) the decay is believed to proceed
dominantly via gluonic penguin processes~\cite{Grossman:1996ke},
and has been evaluated
with generalized factorization 
approaches~\cite{Petrov:1997yf,Ali:1998eb,Du:1998hs}. The measured branching
fractions are, however, significantly larger than these expectations.
This has led to speculations that SU(3)-singlet couplings unique to the
\etap{} meson or new physics~\cite{Xiao:2001uh,Khalil:2003bi} 
contribute to the amplitude. More precise measurements are needed to 
constrain the amplitudes and to distinguish between theoretical models.
Measurements of ratios of branching fractions, which reduce
the effects of form factor uncertainties, are especially
useful in this regard.

Additional constraints can be provided by the direct $CP$ asymmetry,
$\ACP = \frac{\BF(\Bbar\to\bar{f}) - \BF(B\to f)}
          {\BF(\Bbar\to\bar{f}) + \BF(B\to f) },
$
where $f$ is the final state and $\bar{f}$ is its $CP$ conjugate. 
Direct $CP$ violation in the \bteppip{} mode can be large in the 
SM~\cite{Fu:2003fy}
while a non-zero value for \ACP{} in \btepkp{} may require a new
physics contribution~\cite{Grossman:1996ke}.

In this Letter, we report evidence of \bteppi ,
improved measurements of \btepk{} and a direct $CP$ violation search 
in the charged $B$ meson decay modes. The results are based on a data
sample that contains $386 \times 10^6$ \BB{} pairs, i.e. 35 times larger than
our previous dataset~\cite{Abe:2001pf}, 
collected  with the Belle detector at the KEKB asymmetric-energy
$e^+e^-$ (3.5 on 8~GeV) collider~\cite{KEKB}.
KEKB operates at the $\Upsilon(4S)$ resonance 
($\sqrt{s}=10.58$~GeV).

The Belle detector is a large-solid-angle magnetic
spectrometer that
consists of a silicon vertex detector,
a 50-layer central drift chamber (CDC), an array of
aerogel threshold \v{C}erenkov counters (ACC), 
a barrel-like arrangement of time-of-flight
scintillation counters (TOF), and an electromagnetic calorimeter
comprised of CsI(Tl) crystals located inside 
a superconducting solenoid coil that provides a 1.5~T
magnetic field.  An iron flux-return located outside of
the coil is instrumented to detect $K_L^0$ mesons and to identify
muons.  The detector
is described in detail elsewhere~\cite{Belle}.
Two inner detector configurations are used. A 2.0 cm beampipe
and a 3-layer silicon vertex detector is used for the first sample
of $152 \times 10^6$ \BB{} pairs (Set I), while a 1.5 cm beampipe, a 4-layer
silicon detector and a small-cell inner drift chamber are used to record  
the remaining $234 \times 10^6$ \BB{} pairs (Set II)~\cite{Ushiroda}.  

Charged hadrons are identified 
by combining information from the CDC ($dE/dx$),
ACC and TOF systems. Both kaons and pions are selected with an efficiency
of about 86\%. Tighter criteria are applied
to the pion candidate in \bteppip , resulting in an efficiency (kaon
misidentification probability) of about 77\% (4\%). 

The \etap{} mesons are reconstructed via two decay chains:
\epp{} (with $\eta \to \gamma \gamma)$ and \erg .
We reconstruct $\pi^0$, $\rho^0$, $\eta$, $\eta'$ and $K_S^0$ candidates
using the mass windows given in Table~\ref{tab:cuts}. In addition, we require
the following. All photons are required to have an energy of at least 50 MeV,
photons from \etap{} in \erg{} of at least 100 MeV.   
The transverse momenta of $\pi^\pm$ for $\rho^0$
candiates have to be greater than 200 MeV/$c$. The vertex of the
$K_S^0\to\pi^+\pi^-$ has to be displaced from the interaction point
and the $K_S^0$ momentum direction must be consistent with its flight
direction. 
For \bteppio , we
require $|h_{\pi^0}| = \frac{E(\gamma_1)-E(\gamma_2)}{E(\gamma_1)+E(\gamma_2)} 
<0.95 \,(0.6)$ for \epp{} (\erg ),
where $E(\gamma_{1,2})$
is the energy of the two $\pi^0$ decay photons.
Similarly, we require $|h_\eta| < 0.85$.
\begin{table}[htb]
\caption{Mass windows to reconstruct intermediate states.}
\label{tab:cuts}
\begin{tabular}
{@{\hspace{0.5cm}}l@{\hspace{0.5cm}}@{\hspace{0.5cm}}l@{\hspace{0.5cm}}}
\hline \hline
mode	&	mass window (MeV\cs ) \\
\hline
$\pi^0\to\gamma\gamma$	&	[118,150] \, $\pm 2.5 \sigma$	 \\
$\rho^0\to\pi^+\pi^-$  	&	[550,870] \, ---	 \\
$\eta\to\gamma\gamma$  	&	[500,570] \, $+$2.5$\sigma/-$3.3$ \sigma$	 \\
$\epp$ (in $\etap K$) 	&	[945,970] \, $\pm 3.4 \sigma$	 \\
$\erg$ (in $\etap K$) 	&	[935,975] \, $\pm 3 \sigma$	 \\
$\epp$ (in $\etap \pi)$  &	[950,965] \, $\pm 2.5 \sigma$	 \\
$\erg$ (in $\etap \pi)$  &	[941,970] \, $\pm 2.5 \sigma$	 \\
$K_S^0\to\pi^+\pi^-$	& 	[485,510] \, $\pm 3 \sigma$	 \\
\hline \hline
\end{tabular}
\end{table}

$B$ meson candidates are reconstructed combining an $\etap$ meson with 
a pion or kaon candidate. Two kinematic variables are used to extract the 
$B$ meson
signal: the energy difference, $\de = E_B-\ebeam$, 
and the beam-energy constrained
mass, $\mb =\sqrt{\ebeam^2/c^4 - (P_B/c)^2}$, 
where \ebeam{} is the beam energy and $E_B$
and $P_B$ are the reconstructed energy and momentum of the $B$ candidate in the
$\Upsilon(4S)$ rest frame. Events satisfying the requirements $\mb > 5.2$
GeV\cs{} and $|\de| <0.25$ GeV are selected for further analysis.
Around 10\% of these events have multiple B candidates. 
Among these candidates the one with the smallest $\chi^2_{\text{vtx}} 
+ \chi^2_{\etap}$ is selected, where $\chi^2_{\text{vtx}}$ 
is a goodness of vertex fit for all charged particles
and $\chi^2_{\etap} =[(M(\etap)-m_{\etap}/\sigma_{\etap}]^2$,
where $M(\etap)$ is the \etap{} candidate mass, $m_{\etap}$ 
is the nominal mass of the \etap{} and $\sigma_{\etap} = 8$ MeV\cs{}
is the width of the \etap{} mass distribution.

Several event shape variables (defined in the center of mass frame) 
are used to distinguish
the spherical \BB{} topology from 
the jet-like $q\bar q$ continuum events. 
The thrust angle $\theta_T$ is defined
as the angle between the \etap{} momentum direction and 
the thrust axis formed by all particles not belonging to the reconstructed $B$
meson.
Jet-like events tend to peak near $|\cost| = 1$,
while spherical events have a uniform distribution.
The requirement $|\cost|<0.9$ is applied prior
to all other event topology selections.

Additional continuum suppression is obtained by using
modified Fox-Wolfram moments~\cite{SFW} and 
the angle $\theta_B$ between the flight 
direction of the reconstructed $B$ candidate and the beam axis. A 
Fischer discriminant (\FD)~\cite{fisher:1936} is formed from a linear 
combination of $\cost$, $\sperp$ and five modified Fox-Wolfram moments. 
$\sperp$ is the ratio of
the scalar sum of the transverse momenta of all tracks outside a 
$45^{\circ}$ cone around the $\etap$ direction 
to the scalar sum of their total momenta.
These variables are then combined to form an event-topology likelihood function
$\LK_s$ ($\LK_{q\bar{q}}$), 
where $s$ ($q\bar{q}$) represents signal (continuum background).
We include the quality of
the $B$ flavor tagging of the accompanying $B$ meson to improve continuum
rejection. The
standard Belle $B$ tagging package~\cite{TaggingNIM} is used, 
which gives the $B$ flavor
and a tagging quality $r$ ranging from zero for no flavor to unity
for unambiguous flavor assignment. The data is divided into three
$r$ regions.
Signal-like events are selected by 
applying likelihood ratio $\LR = \LK_{s}/(\LK_{s} + \LK_{q\bar q})$ requirements
optimized on MC events in the three $r$ regions separately.
For channels with an \erg{} decay an additional
variable $\theta_{\hel}$, which is the angle between 
the $\etap$ momentum and the direction of one of
the decay pions in the $\rho^0$ rest frame, is
included for better signal-background separation. 

The branching fractions are extracted using extended unbinned 
maximum-likelihood fits to two-dimensional (\de ,\mb ) distributions for the
\epp{} and \erg{} sub-decays simultaneously.
The extended likelihood function used is:
\begin{eqnarray}
L(N_S,N_{B_j}) &=&  \frac{e^{-(N_S+\sum_jN_{B_j})}}{N!} 
\prod_{i=1}^{N}\biggl[N_{S} P_S(\de_i,M_{\mbox{\scriptsize bc}_i})  \nonumber \\
& & + \sum_j N_{B_j} P_{B_j}(\de_i,M_{\mbox{\scriptsize bc}_i})\biggr]
\label{eq:ns-lkhd}
\end{eqnarray}
where $N_S$ ($N_{B_j}$) is the number of signal events (background events of
source $j$) with probability density functions (PDFs) 
$P_S$ ($N_{B_j}$) and the index 
$i$ runs over the total number of events $N = N_S + \sum_j N_{B_j}$.

The reconstruction efficiencies are determined from signal MC samples, 
using the EvtGen package~\cite{bib:Evtgen} with final
state radiation simulated by the PHOTOS package~\cite{bib:Photos}.
The efficiencies are calculated separately for
both Set I and Set II. The absolute
efficiency for Set II is typically about 0.5\%
larger than for Set I (for efficiencies averaged over the two sets 
see Table~\ref{tab:ohwell}).
The signal yield is expressed as
$
N_S =  \eff_1 \, N_{(\BB)_1} \, \BF +  \eff_2 \, N_{(\BB)_2} \, \BF ,
$
where $\BF$ is the signal branching fraction, and $\epsilon_i$ and
$N_{(\BB)_i}$ are the efficiency and the number of \BB{} pairs
for Set I and Set II. The numbers of $B^+B^-$ and $B^0\Bbaro$
pairs are assumed to be equal.
Correction factors due to differences
between data and MC are included for the charged track
identification and photon, $\pi^0$ and $\eta$ reconstruction, 
resulting in an overall correction factor of $\approx 0.9$

The PDF shapes for each contribution are determined by MC studies.
The signal shapes for \de{} and \mb{} are assumed to be independent. 
We model the signal using a Gaussian with an
exponential tail (Crystal Ball function)~\cite{CBline} 
plus a Gaussian for \de{} and a Gaussian with an exponential
tail for \mb. 

We consider four types of backgrounds separately in the fit:
continuum, $b\to c$ and two types of charmless decays. 
Continuum background is modeled by a first or second
order polynomial for \de{} and an ARGUS function~\cite{bib:ARGUS} 
for \mb. Charmless
$B$ decays and $b\to c$ backgrounds are modeled with smoothed two-dimensional  
histograms. 
The contributions from charmless $B$ decays are split into two components,
one for the decay with the largest contribution and one for all
other charmless decays.
For \btepkp , the dominant mode,  which is modeled
separately is $B\to \etap K^*$;
for \btepks{} it is $B\to \rho^0 K_S^0$; for \bteppio{} it is $B\to \rho \rho$;
and for \bteppip{} it is the \btepkp{} feed-down.
The feed-down in \bteppip{} is modeled 
with the same PDFs as used for the $\etap K^+$ signal, 
shifted and with a corrected 
width in \de .

\begin{figure*}[!htb]
\unitlength1.0cm
\centerline{
\epsfxsize 1.6 truein \epsfbox{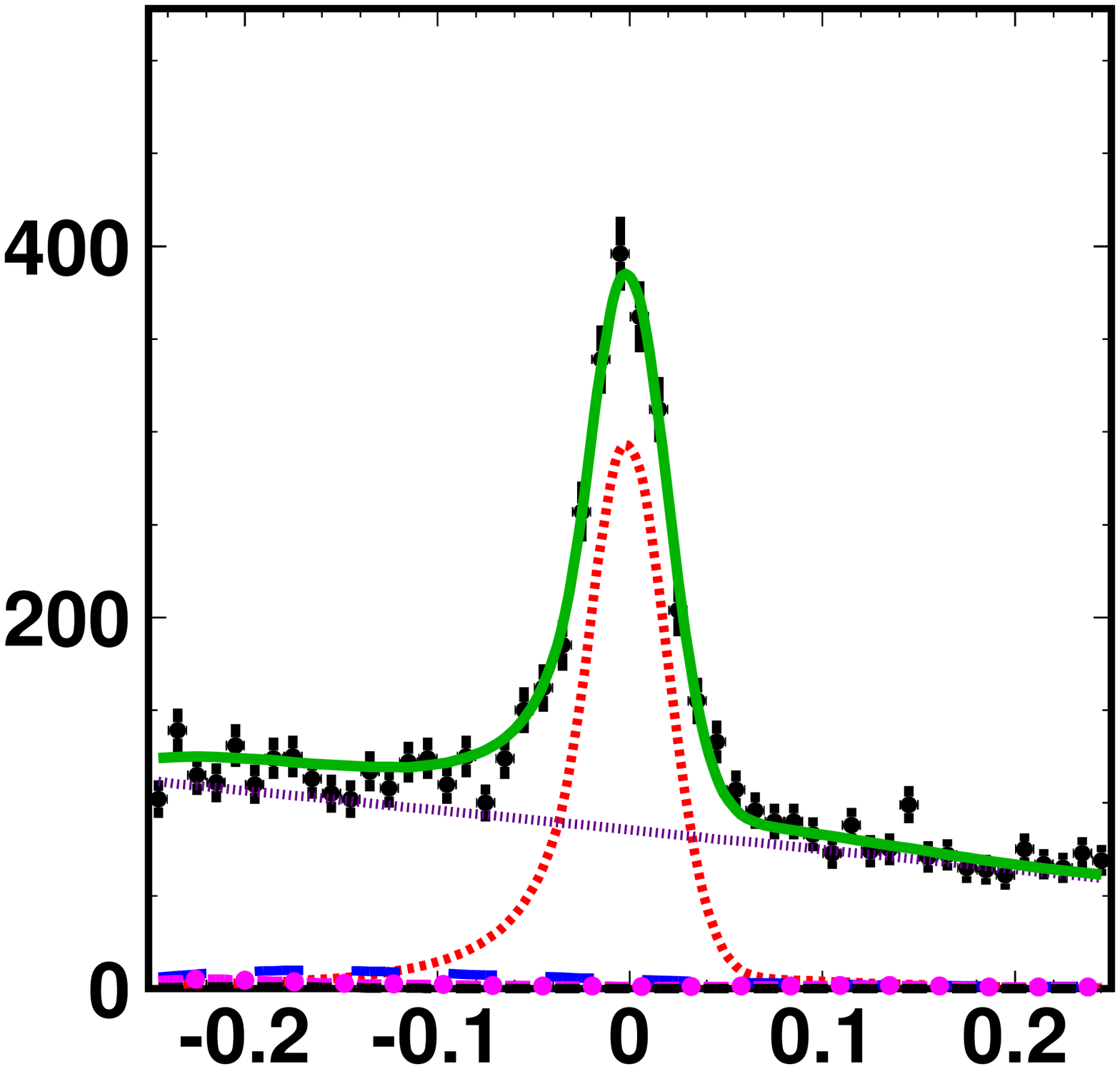}
\epsfxsize 1.6 truein \epsfbox{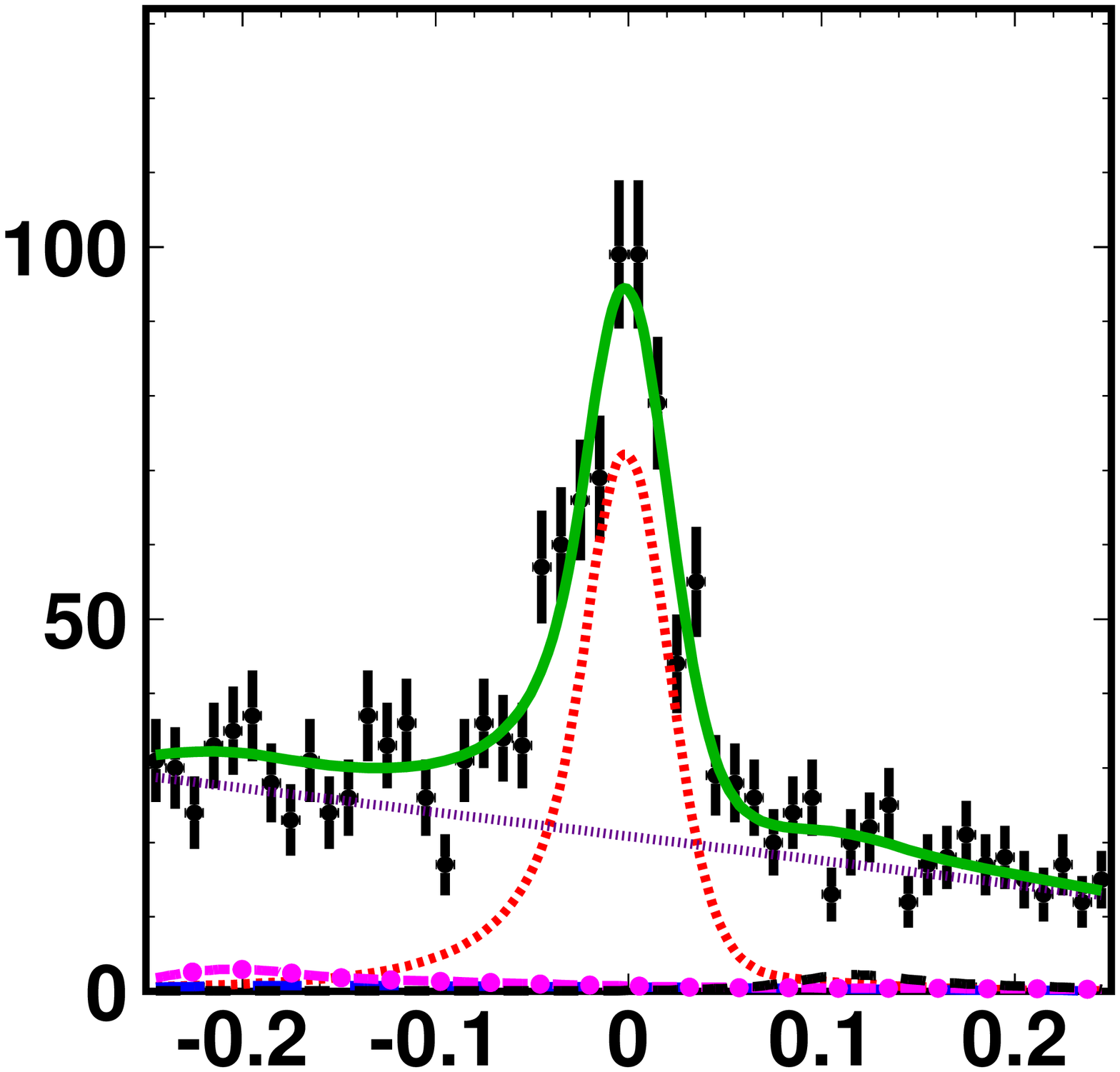}
\epsfxsize 1.6 truein \epsfbox{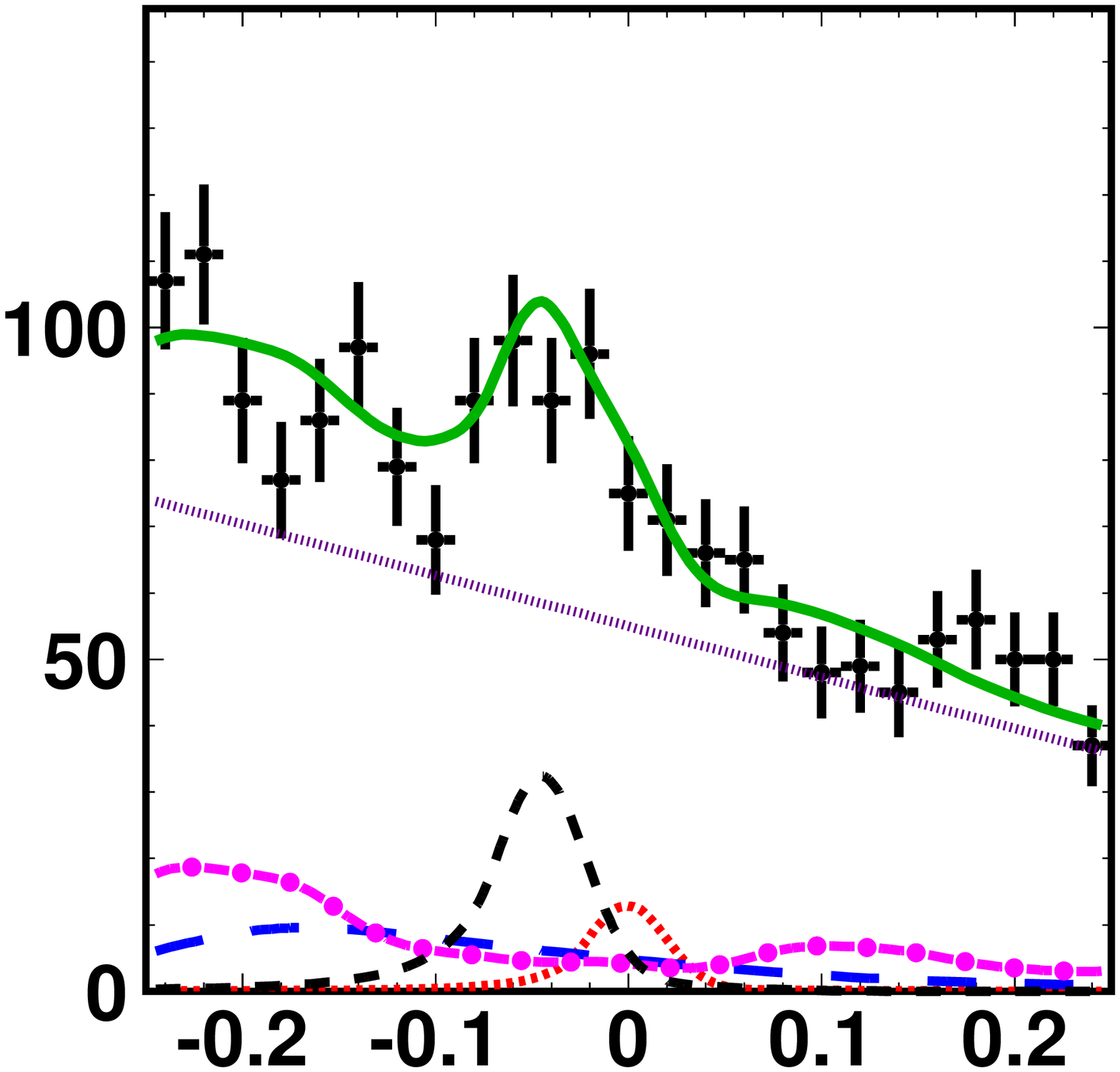}
\epsfxsize 1.6 truein \epsfbox{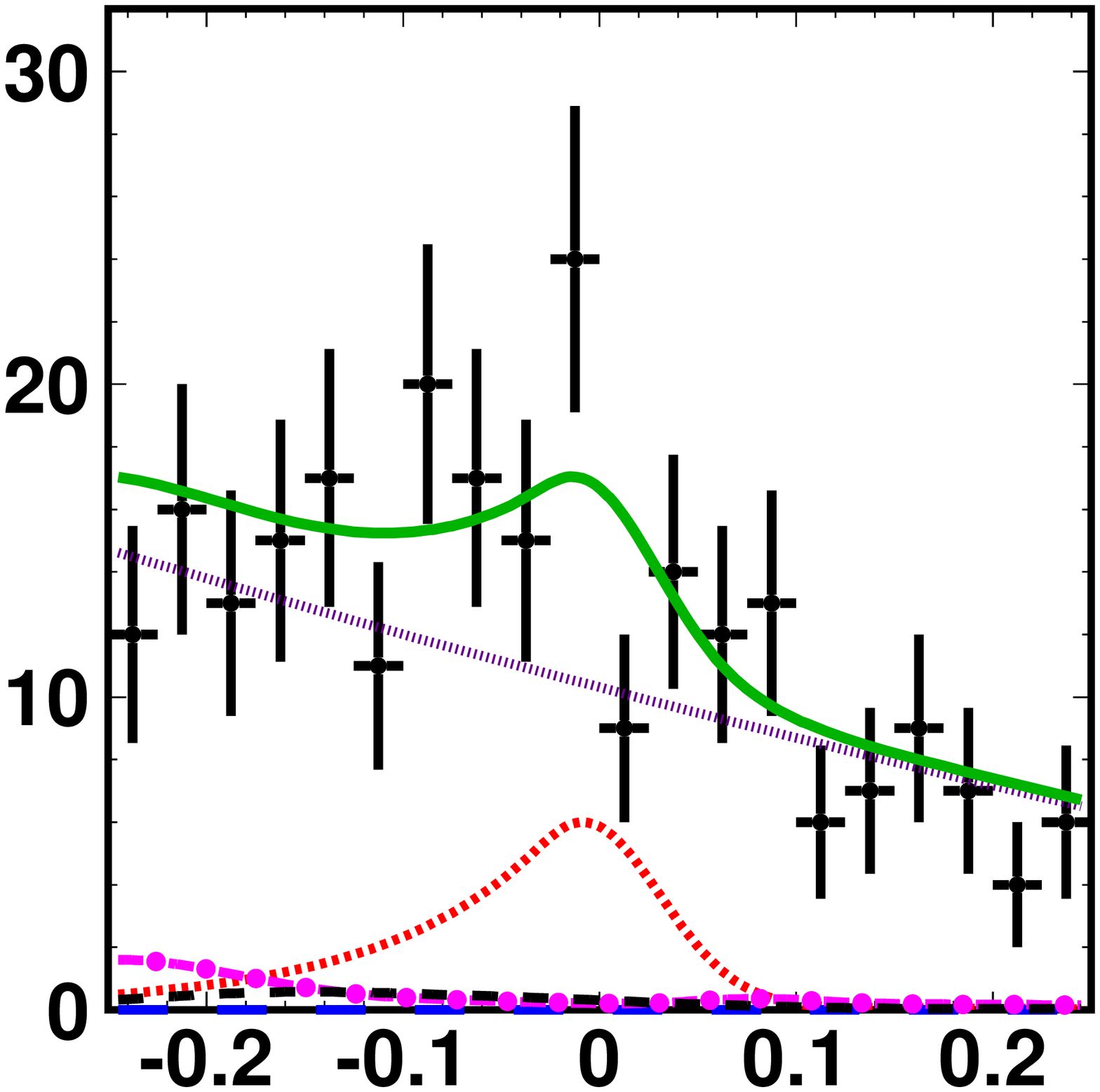}
\put(-2.6,0.0){\footnotesize{\sf\shortstack[c]{\de{} [GeV]}}}
\put(-6.7,0.0){\footnotesize{\sf\shortstack[c]{\de{} [GeV]}}}
\put(-10.9,0.0){\footnotesize{\sf\shortstack[c]{\de{} [GeV]}}}
\put(-15.1,0.0){\footnotesize{\sf\shortstack[c]{\de{} [GeV]}}}
\put(-3.3,3.45){\footnotesize{\sf\shortstack[c]{d) \, \bteppio}}}
\put(-7.5,3.45){\footnotesize{\sf\shortstack[c]{c) \, \bteppip}}}
\put(-11.65,3.45){\footnotesize{\sf\shortstack[c]{b) \, \btepks}}}
\put(-15.85,3.45){\footnotesize{\sf\shortstack[c]{a) \, \btepkp}}}
\put(-16.85,1.2){\footnotesize{\sf\shortstack[c]{\rotatebox{90}{Events / 10 MeV}}}}
\put(-12.6,1.2){\footnotesize{\sf\shortstack[c]{\rotatebox{90}{Events / 10 MeV}}}}
\put(-8.45,1.2){\footnotesize{\sf\shortstack[c]{\rotatebox{90}{Events / 20 MeV}}}}
\put(-4.2,1.2){\footnotesize{\sf\shortstack[c]{\rotatebox{90}{Events / 25 MeV}}}}
}
\centerline{
\epsfxsize 1.6 truein \epsfbox{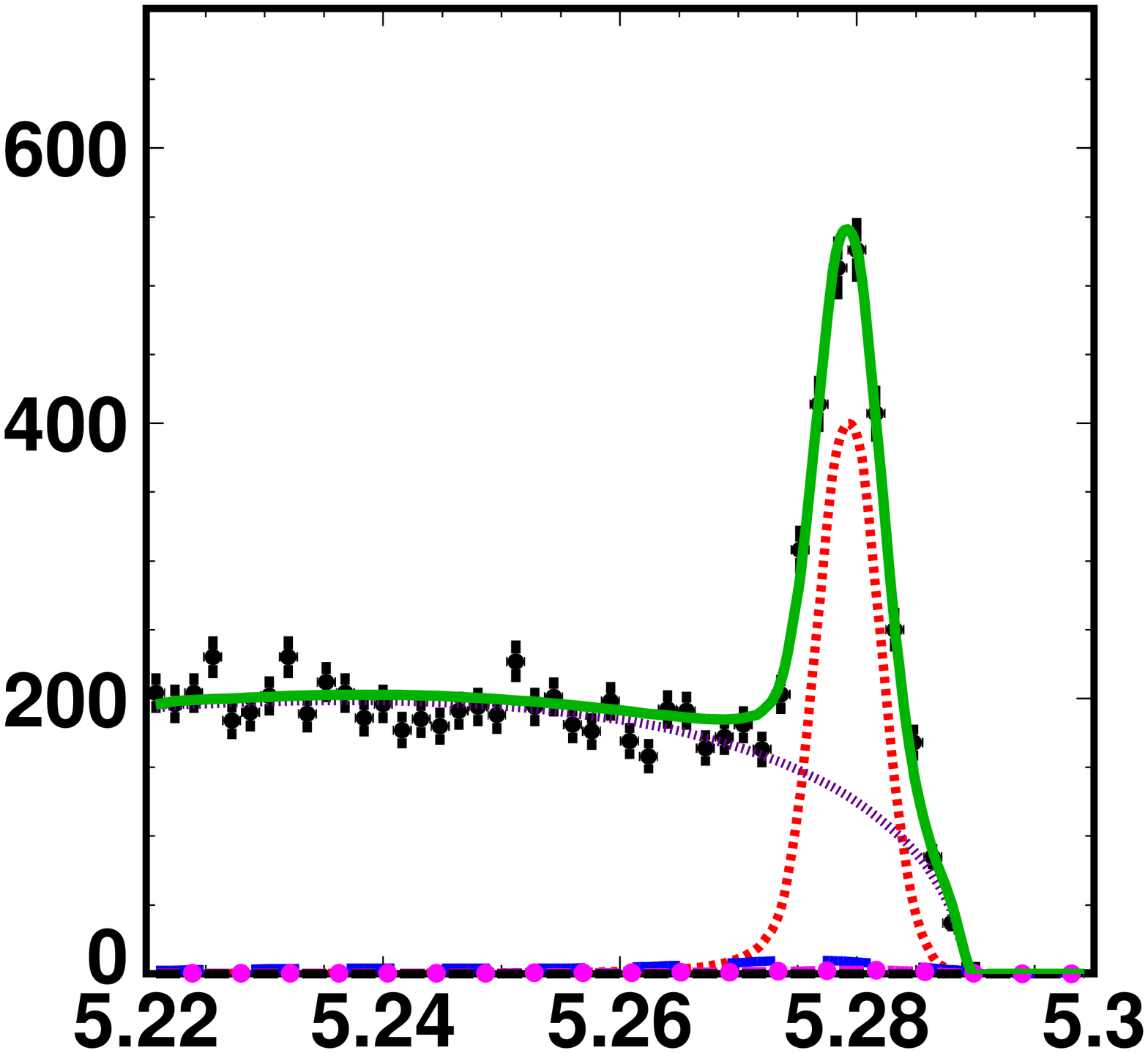}
\epsfxsize 1.6 truein \epsfbox{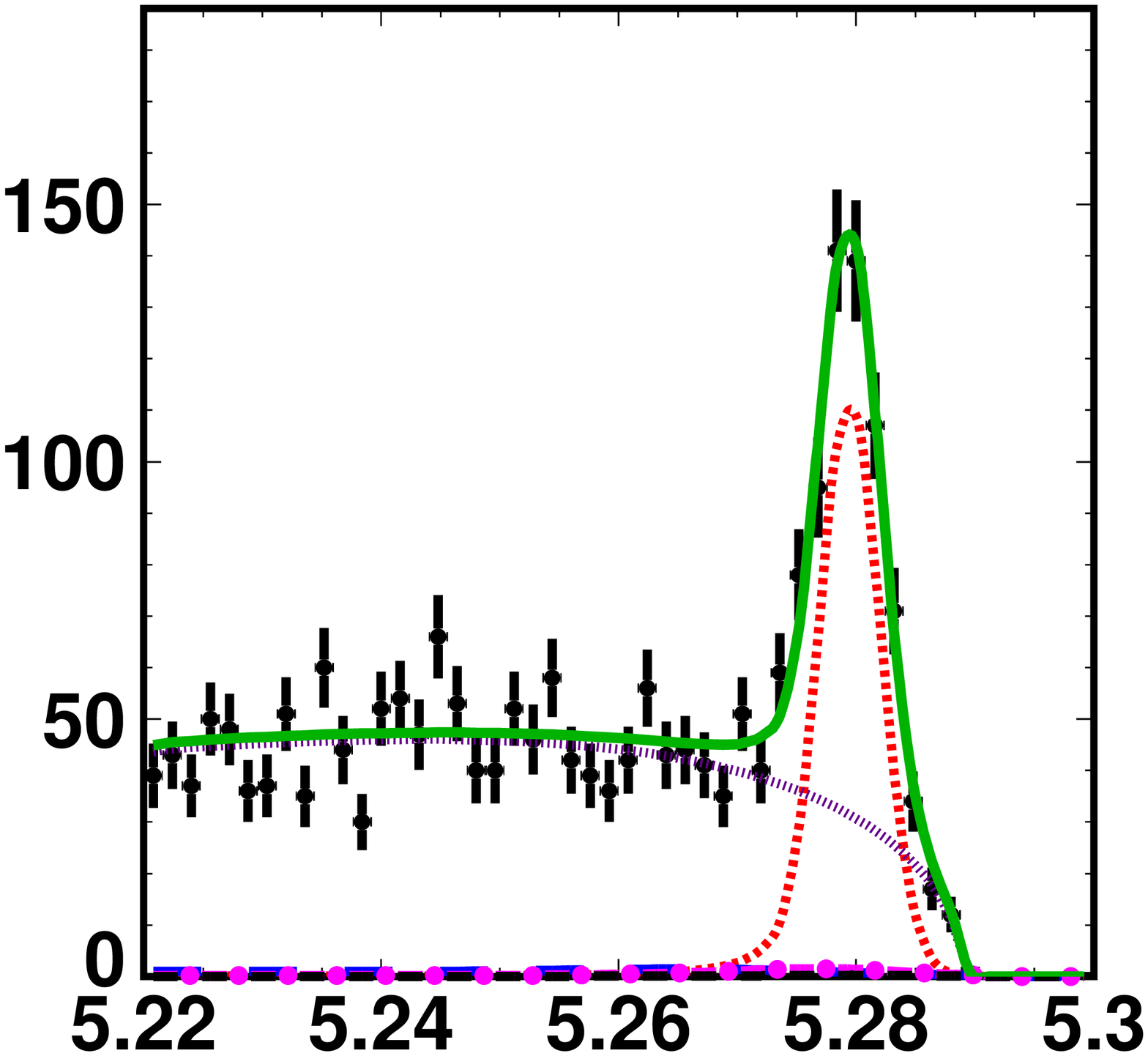}
\epsfxsize 1.6 truein \epsfbox{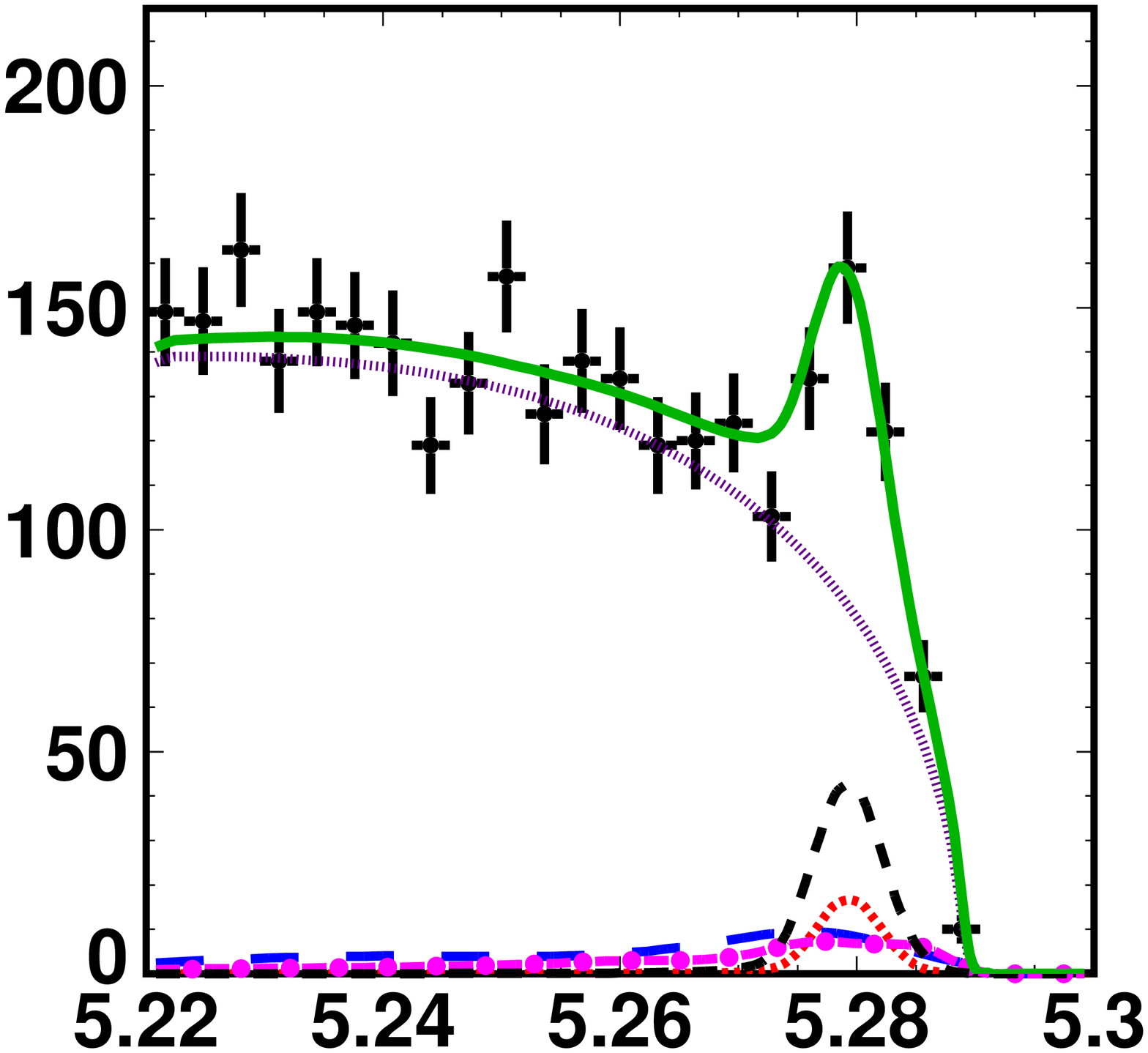}
\epsfxsize 1.6 truein \epsfbox{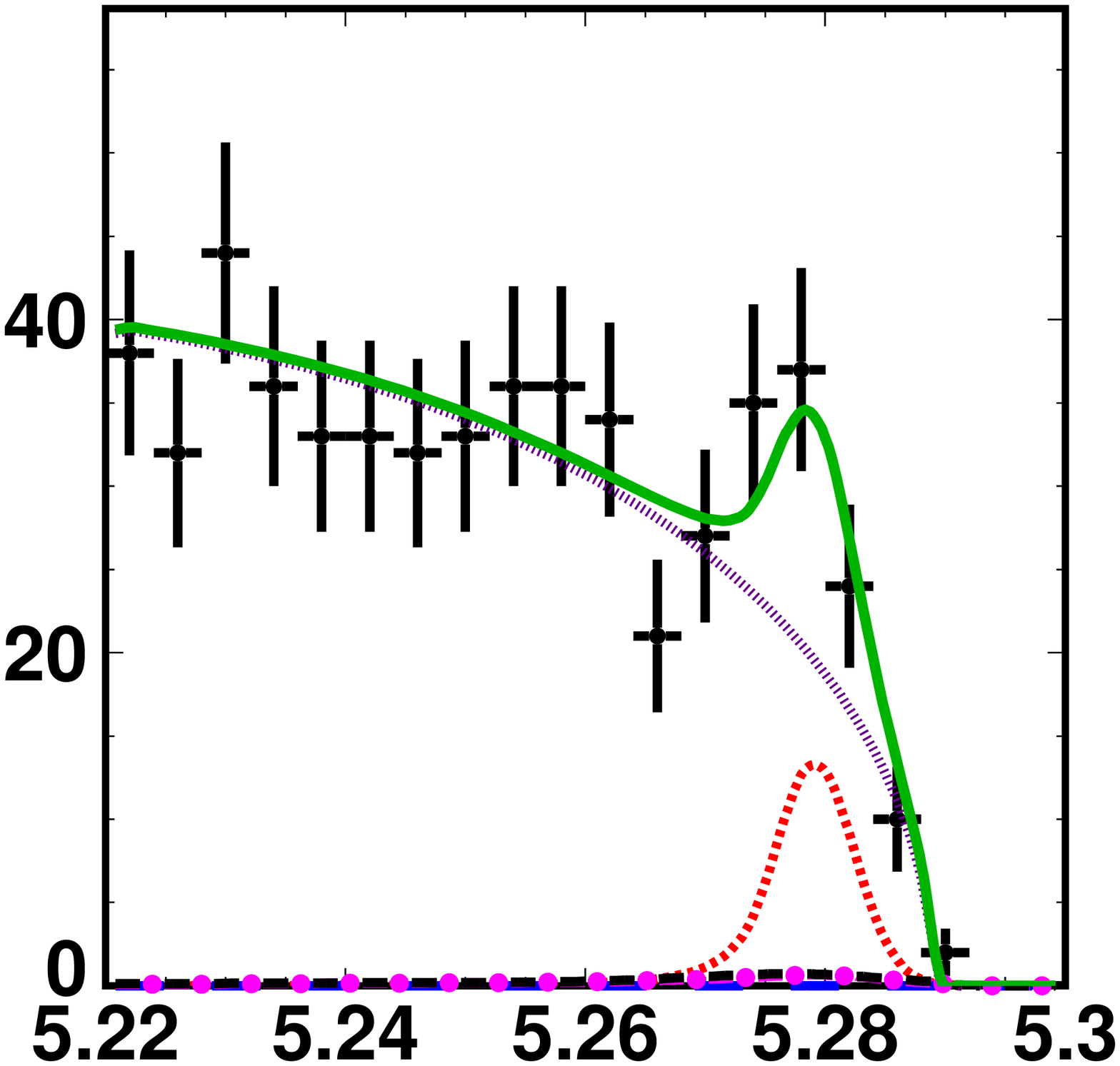}
\put(-2.6,0.0){\footnotesize{\sf\shortstack[c]{\mb{} [GeV\cs ]}}}
\put(-6.8,0.0){\footnotesize{\sf\shortstack[c]{\mb{} [GeV\cs ]}}}
\put(-11.0,0.0){\footnotesize{\sf\shortstack[c]{\mb{} [GeV\cs ]}}}
\put(-15.2,0.0){\footnotesize{\sf\shortstack[c]{\mb{} [GeV\cs ]}}}
\put(-3.3,3.45){\footnotesize{\sf\shortstack[c]{h)}}}
\put(-7.5,3.45){\footnotesize{\sf\shortstack[c]{g)}}}
\put(-11.65,3.45){\footnotesize{\sf\shortstack[c]{f)}}}
\put(-15.85,3.45){\footnotesize{\sf\shortstack[c]{e)}}}
\put(-15.85,3.1){\footnotesize{\sf\shortstack[c]{\, +  \, data}}}
\put(-15.85,2.85){\footnotesize{\sf\shortstack[c]{\textcolor{red}{ - - -  signal}}}}
\put(-15.85,2.6){\footnotesize{\sf\shortstack[c]{\textcolor{violet}{ $ \cdot \cdot \cdot \cdot $ continuum}}}}
\put(-15.85,2.35){\footnotesize{\sf\shortstack[c]{\textcolor{blue}{ $-\, -$  $b\to c$ bg.}}}}
\put(-15.85,2.1){\footnotesize{\sf\shortstack[c]{\textcolor{magenta}{ --$\bullet$--  rare $B$}}}}
\put(-15.85,1.85){\footnotesize{\sf\shortstack[c]{\textcolor{green}{ -----  combined}}}}
\put(-16.85,1.2){\footnotesize{\sf\shortstack[c]{\rotatebox{90}{Events / 2 MeV\cs}}}}
\put(-12.6,1.2){\footnotesize{\sf\shortstack[c]{\rotatebox{90}{Events / 2 MeV\cs}}}}
\put(-8.45,1.2){\footnotesize{\sf\shortstack[c]{\rotatebox{90}{Events / 4 MeV\cs}}}}
\put(-4.2,1.2){\footnotesize{\sf\shortstack[c]{\rotatebox{90}{Events / 5 MeV\cs}}}}
\put(-7.5,1.3){\footnotesize{\sf\shortstack[c]{-- -- \, \etap $K^+$}}}
}
\caption{\label{fig:K} \de{} and \mb{} distributions for \btepkp{} (a, e),
\btepks{} (b, f), \bteppip{} (c, g) and \bteppio{} (d, h), respectively,
for the
region $\mb>5.27$ GeV\cs and $-0.12$ GeV $<\de<0.08$ GeV for \bteppio{} and 
$-0.1$ GeV $<\de<0.06$ GeV for all others.}
\end{figure*}

The continuum shape parameters are allowed to float in all modes.
The signal mean and width parameters are free
for the kaonic modes. For the \bteppip{} mode
these parameters are fixed to the values
obtained from the charged kaon mode.
For \epp{} modes our background MC studies show that no contributions from 
$b\to c$ decays are expected. 
The sizes of background contributions other than continuum are constrained to 
the values expected from MC studies. The \btepkp{}
component in the \bteppip{} decay is fixed by the branching fraction of
\btepkp{} as measured here and the probability of kaons faking pions.
A simultaneous fit with the branching fraction
and the charge asymmetry as common parameters is used. 
The resulting projection plots are shown in
Fig.~\ref{fig:K}. 
The reconstruction efficiencies and fit results are given in 
Table~\ref{tab:ohwell}. 

We find first evidence for the neutral decay:
$$
\BF(B^{0}\to \etap \pi^0) = [ 2.79 ^{+1.02}_{-0.96}\text{(stat)}
^{+0.25}_{-0.34}\text{(syst)}]\times 10^{-6},
$$ and evidence for the charged decay:
$$
\BF(B^{\pm}\to \etap \pi^{+}) = [ 1.76 ^{+0.67}_{-0.62}\text{(stat)}
^{+0.15}_{-0.14}\text{(syst)}]\times 10^{-6}.
$$
The ratio of the branching fractions for
charged and neutral \btepk{} decays is found to be $1.17 \pm 0.08\pm 0.03$, 
assuming equal production of $B^+B^-$ and $B^0\Bbaro$.
The charge asymmetries for the \bteppip{} and \btepkp{} decay modes,
listed in Table~\ref{tab:ohwell}, show no significant deviation from zero.

Systematic errors are estimated with various high statistics data samples. 
The dominant sources are the uncertainties 
of the reconstruction efficiency of charged tracks (3--4\%), 
the uncertainties in the
reconstruction efficiencies for $\eta$ mesons and photons (3--6\%) and 
the uncertainty of the PDF shapes and parameters (1--9\%).
Other systematic uncertainties arise from the $K$ feed-down in \bteppip{} 
($\approx 2$\%), 
the differences between data and MC for \de{} and \mb{} in \bteppip{} 
($\approx 4$\%),
the $K_S$ reconstruction efficiency uncertainty (4\%), 
the uncertainty of the sub-decay branching fractions as
given by the Particle Data Group (PDG)~\cite{bib:PDG04} (1.5\%), 
the number of \BB{} mesons produced (1\%), the efficiency differences
due to signal simulation by different MC generators (1.4\%), 
the uncertainty in
the efficiency (1\%) and the uncertainty from particle identification (0.7\%). 
The errors are added in quadrature and we find the systematic
errors to be $\pm5.4$\%, $\pm7.3$\%, $^{+8.5}_{-7.7}$\% and
$^{+8.8}_{-12.1}$\% for \btepkp, \btepks, \bteppip{} and \bteppio{} decays,
respectively.
For the charge asymmetry, efficiency based systematic errors cancel 
out.  We estimate the possible detector bias on
\ACP{} from the charge asymmetry of the continuum background in the
\btepkp{} sample which is obtained
simultaneously from the fit.  We assign 0.02 as systematic
error both for \btepkp{} and $\etap \pi^+$.
Other contributions from fitting and
normalization together result in a systematic error of 0.003 for \btepkp. For
\bteppip{} the uncertainties from PDF shapes and feed-down contributions add up
to $^{+0.03}_{-0.04}$.

The significance of the \bteppip{} yield is $3.2\sigma$,
which is calculated as $\sigma = \sqrt{2\ln({\cal L}_{\rm max}/{\cal L}_{0})}$,
where ${\cal L}_{\rm max}$ and ${\cal L}_{0}$ denote
the maximum likelihood value and the
likelihood value at zero branching fraction, respectively.
The systematic error
is included in the significance calculation.
For \bteppio{} the corresponding significance with systematics is $3.1\sigma$.
\begin{table*}[htb]
\caption{Signal efficiencies ($\epsilon_{\text{tot}}$) with
sub-decay branching fractions included and averaged for Set I 
and Set II for \epp{}
and \erg , total signal yields $N_S$, branching fractions \BF , signal
asymmetries \ACP ,
and significances $\sigma$.
The first errors are statistical and
the second (if given) are systematic errors.}
\label{tab:ohwell}
\begin{tabular}
{@{\hspace{0.5cm}}l@{\hspace{0.5cm}}||@{\hspace{0.5cm}}c@{\hspace{0.5cm}}
@{\hspace{0.5cm}}c@{\hspace{0.5cm}}@{\hspace{0.5cm}}c@{\hspace{0.5cm}}
@{\hspace{0.5cm}}c@{\hspace{0.5cm}}}
\hline \hline
	&	\btepkp		& \btepks 	& \bteppip	& \bteppio \\
\hline
$\epsilon_{\text{tot}}(\epp)$ [\%] 
	& $4.31\pm 0.03$ & $1.19\pm 0.03$ & $2.84\pm 0.03$ & $1.72\pm 0.02$ \\
$\epsilon_{\text{tot}}(\erg)$ [\%] 
	& $2.78 \pm 0.04$ & $1.07\pm 0.04$ & $2.89\pm 0.04$ & $1.72\pm 0.03$ \\
$N_S$		
	& $1895.7\pm 59.5$ & $515.3\pm 31.7$ & $39.0\pm 13.2$ & $35.8\pm 12.7$\\
\BF [$10^{-6}]$	      
	& $69.2{}\pm 2.2{}\pm 3.7$ & $58.9^{+3.6}_{-3.5}{}\pm 4.3$ 
		& $1.76^{+0.67}_{-0.62} {}^{+0.15}_{-0.14}$ 
		& $2.79 ^{+1.02}_{-0.96} {}^{+0.25}_{-0.34}$ \\
\ACP	& $0.028\pm 0.028\pm 0.021$ &
			   --- & $0.20^{+0.37}_{-0.36}{} \pm 0.04$ & --- \\
$\sigma$		& $>10$ & $>10$ & 3.2 & 3.1 \\ 
\hline \hline
\end{tabular}
\end{table*}

In summary, evidence for \bteppi{} with greater than $3 \sigma$
significance is found and improved measurements for the charged and 
neutral \btepk{} decays are reported. 
The measured branching fractions of \btepk{} decays supersede our
previous results~\cite{Abe:2001pf} and are consistent
with the measurements of CLEO~\cite{Richichi:1999kj} and 
BaBar~\cite{Aubert:2005bq}.
No charge asymmetry is observed in
the decay modes \btepkp{} and \bteppip. 

We thank the KEKB group for excellent operation of the
accelerator, the KEK cryogenics group for efficient solenoid
operations, and the KEK computer group and
the NII for valuable computing and Super-SINET network
support.  We acknowledge support from MEXT and JSPS (Japan);
ARC and DEST (Australia); NSFC and KIP of CAS 
(contract No.~10575109 and IHEP-U-503, China); DST (India); 
the BK21 program of MOEHRD, and the
CHEP SRC and BR (grant No. R01-2005-000-10089-0) programs of
KOSEF (Korea); KBN (contract No.~2P03B 01324, Poland); MIST
(Russia); ARRS (Slovenia);  SNSF (Switzerland); NSC and MOE
(Taiwan); and DOE (USA).


\begin{thebibliography}{99}
\bibitem{CC}
Throughout this paper, 
the inclusion of the charge conjugate mode decay is implied
unless otherwise stated.
\end{thebibliography}
\end{document}